\documentclass{PoS}
\usepackage[authoryear,square]{natbib}
\bibpunct{(}{)}{;}{a}{}{,}

\title{Detecting the cosmic web with radio surveys}
\ShortTitle{Cosmic web in radio surveys}
\author{\speaker{Franco Vazza}$^{1}$, 
              Marcus Br\"{u}ggen$^1$,
              Claudio Gheller$^3$,
              Chiara Ferrari$^2$,
              Annalisa Bonafede$^1$
\\
	$^1$Hamburg University (Hamburg Observatory), Gojenbergsweg 112, 21029,
Germany;\\
        $^2$Laboratoire Lagrange, UCA, OCA, CNRS, Blvd de l'Observatoire, CS 34229, 06304 Nice cedex 4, France;\\
       $^3$ETHZ-CSCS, Via Trevano 131, CH-6900 Lugano, Switzerland;
   
    E-mail:\email{franco.vazza@hs.uni-hamburg.de}
 }

\abstract{{\bf Abstract}  We study the challenges to detect the cosmic web at
radio wavelengths with state-of-the-art cosmological simulations of
extragalactic magnetic fields. The incoming generation of radio surveys operating at low frequency, like
LOFAR, SKA-LOW and MWA will have the best chance to detect the large-scale, low surface brightness emission from the shocked cosmic web.
 The detected radio emission  will enable to constrain the average magnetisation level of the gas in filaments and the 
 acceleration efficiency of electrons by strong  shocks. In case of detections, through statistical modelling (e.g. correlation functions)  it  will be possible  to discriminate  among competing scenarios for the magnetisation of large-scale structures (i.e. astrophysical versus primordial scenarios), making radio surveys an important probe of cosmic magnetogenesis.}

\FullConference{EXTRA-RADSUR2015 (*)\\
 20--23 October 2105 \\
 Bologna, Italy
  \bigskip
   \hrule
   \bigskip

 \textnormal{(*) This conference has been organised 
 with the support of the Ministry of Foreign Affairs
 and International Cooperation, Directorate General for 
 the Country Promotion (Bilateral Grant Agreement 
 ZA14GR02 - Mapping the Universe on the Pathway to SKA)}
 }

\newcommand{\enzo}{\it{\small ENZO}}

\begin{document}

\section{Introduction}

Deep radio surveys at low frequencies might be able to detect the cosmic web 
in radio continuum, and start imaging the shocked warm-hot intergalactic medium (WHIM), that 
contains  half of baryons in the universe \citep[e.g.][]{2001ApJ...552..473D,gheller15}.
This should be possible because the gas in the cosmic web is expected to be surrounded by strong accretion shocks, where the cosmic gas is first shock-heated \citep[][]{1972A&A....20..189S} and particles can be accelerated to relativistic energies, via diffusive shock acceleration \citep[][]{be87}. Owing to their rather large Mach number, $\mathcal{M} \sim 10-10^2$, cosmological accretion shocks should be efficient accelerators of 
both cosmic ray protons \citep[e.g.][]{ry03,pf06} and electrons \citep[e.g.][]{hb07,sk11}.  Relativistic electrons accelerated here might emit synchrotron emission and illuminate the radio cosmic web \citep[][]{2004ApJ...617..281K,2011JApA...32..577B}, provided that the magnetisation of the WHIM is large enough.  
Detecting this emission and using it to infer the level of extragalactic magnetic fields is  
important to study the origin of cosmic magnetism, whose origin is still
debated \citep[e.g.][]{wi11}. Two main scenarios can be identified: turbulent amplification and compression of weak fields generated during phase transitions (hereafter "primordial" scenarios), or magnetisation by galactic winds and outflows powered by star formation feedback, supernovae and AGN (hereafter "astrophysical" scenarios). 
While the magnetic field configuration in clusters at the present epoch should be fairly independent of the initial seed field \citep[][]{donn09,ch14}, the magnetic fields in filaments should carry memory of the initial field \citep[][]{va14mhd,2015MNRAS.453.3999M}. \\
The predictions for the WHIM magnetisation are still very uncertain in MHD simulations, due to 
numerical limitation in resolution and physical uncertainties on the seeding processes. However, cosmological simulations can be used to produce observable predictions for different models, to be tested by radio observations.
In recent work, we produced mock radio observations of the cosmic web and quantified how radio observations can be used to constrain the combination of the acceleration efficiency and of the magnetic fields in the cosmic web \citep[][]{va15ska,va15radio}. In this contribution we review the most important results of this study and include the latest developments on this research.

\section{Methods}

\subsection{Cosmological MHD Simulations}
\label{models}
The simulations we present in this work belong to the CHRONOS++  suite of cosmological magneto-hydrodynamical (MHD) simulations with {\enzo}  \citep[][]{enzo14}  running on the Piz-Daint Supercluster at CSCS (Lugano) and tailored to explore the observable properties of competing seeding scenarios for extragalactic magnetic fields. \\
Here we analyse two re-simulations of a $50^3 \rm Mpc^3$  comoving volume from $z=33$ until $z=0$, with $1200^3$ cells and dark matter particles (for a fixed comoving spatial and mass resolution of  $40 ~\rm kpc$ and $\approx 1.1 \cdot 10^{7} M_{\odot}$). 
A first non-radiative run includes gravity, cosmological expansion and magnetic fields, evolved starting from a uniform primordial seed field of $B_0=10^{-9} \rm G$  comoving (hereafter "primordial'' model). This run is contrasted with a second run (hereafter "astrophysical'' model) that also includes radiative gas cooling, thermal and magnetic feedback from active galactic nuclei (from $z=4$), and assumes a $100$ times lower primordial seed field,  $B_0=10^{-11} \rm G$. 
As in other works \citep[][]{va13feedback} we include AGN feedback by releasing at run-time extra thermal/magnetic energy from high density peaks with halos, $n \geq 10^{-2} \rm particles/cm^3$, with a fixed energy per event ($E_{\rm AGN} = 10^{59} \rm erg$ for the thermal energy and $E_{\rm AGN,B} = 10^{-2} E_{\rm AGN}$ for the magnetic energy, released as a dipole). To evolve MHD equations we use the Dedner cleaning method implemented in {\enzo}, ported onto the GPU by \citet{wang10}.

 \subsection{Radio emission and mock observations}
 \label{mock}
We estimate the continuum radio synchrotron emission by relativistic 
electron accelerated by cosmological shocks, in the diffusive shock acceleration (DSA) scenario, following \citet{hb07}. DSA is assumed to generate supra-thermal electrons that follow a power-law in energy in the downstream region of shocks, and the total radio emission in the downstream region is the convolution from all the contributions of power-law distributions, which have aged as a function of distance from the shock because of synchrotron and Inverse Compton losses.   First, we identify schocks in the simulation with a velocity-based approach \citep[][]{va09shocks} and then we apply the formalism by \citet{hb07}  to compute the radio emission as a function of gas parameters in the cells (density, temperature, magnetic fields). The acceleration efficiency of electrons by shocks, $\xi_e(\mathcal{M},T)$ is a function tailored to statistically reproduce radio emission from cluster shocks, i.e. $\xi_e \sim 10^{-6}-10^{-5}$ for weak merger shocks in the ICM. 
In previous work \citep[][]{va15radio} we estimated that in this scenario, the average radio emission from the shocked WHIM averaged over large comic volumes should reduce to: 
\begin{equation}
\label{eq1}
P_{\rm WHIM}(\nu) \sim 5 \times 10^{-3} \rm \frac{Jy}{deg^2}\frac{100 MHz}{\nu} \cdot \frac{\epsilon_{\rm B}}{0.01} \cdot \frac{\xi_e}{10^{-3}} \ ,
\end{equation}
where $\epsilon_{\rm B}$ is ratio between the magnetic and the thermal energy of the WHIM and $\xi_e$ is the acceleration efficiency of electrons at strong,  $\mathcal{M} \gg 10$ shocks. \\

Our post-processing strategy to include all most salient features of radio continuum imaging has been first discussed in \citet[][]{va15ska} as an application to future SKA observations, and in \citet{va15radio} in connection to several other existing and incoming radio surveys. 
In summary: 1) we compute the radio emission in the reference frame of each simulated volume, and produce maps of total emission by summing the contribution of all cells along the line of sight. 2) The emission is converted into the physical frame of the observer, i.e. we dim the emission for the luminosity distance and include further cosmological dimming. 3) The maps are FFT-transformed to remove the frequencies below the minimum antenna baseline of each specific radio configuration, i.e. we mimic the loss of signal from scales larger than those sampled by the minimum instrumental baseline. This is particularly relevant for the large-scale diffuse WHIM emission. 4)  The maps are converted back into real space and the emission is convolved for the resolution beam with a Gaussian filter. 5) Only where the emission is $\geq 3 \sigma_{\rm rms}$ (where $\sigma_{\rm rms}$ is largest between the thermal or the confusion noise of each instrument)  we consider the simulated emission as detectable.
6) As a simplifying working assumption, we assume a perfect removal of the Milky Way foreground and of all resolved point-like radio sources, as well as  an ideal calibration and deconvolution of the radio data. 

 A public repository of radio maps for the full volumes studied in \citet{va15radio} is available at {\it http://cosmosimfrazza.myfreesites.net/radio-web }.

\section{Results}

\subsection{Overview of previous results}
\label{old}
We studied the detectability of the cosmic web in the primordial scenario of extragalactic magnetic fields in \citet{va15radio}, with the same techniques introduced in Sec.\ref{mock}.
We found that radio surveys can detect the cosmic web if the WHIM is on average amplified at the level of a few $\%$ of the thermal energy. This amplification level is not observed in direct simulations \citep[][]{va14mhd}, yet it would be consistent with the magnetic field detected along at filamentary accretion onto the Coma cluster \citep[][]{bo13} and can be achieved via small-scale dynamo amplification below the resolution limits of existing simulations. It may be as well reached by the fast growing instabilities in a high plasma $\beta$ \citep[e.g.][]{2014MNRAS.440.3226M}. \\
In this scenario, low-frequency ($\leq 300$ MHz) observations have the best chances of a detection, owing to their better sampling of the $\sim $degrees scales traced by the low redshift cosmic web. A typical surface brightness of  $\sim  \rm \mu Jy/arcsec^2$ must be reached at these scales to get a detection.

\begin{figure*}
 \includegraphics[width=0.9\textwidth,height=0.4\textwidth]{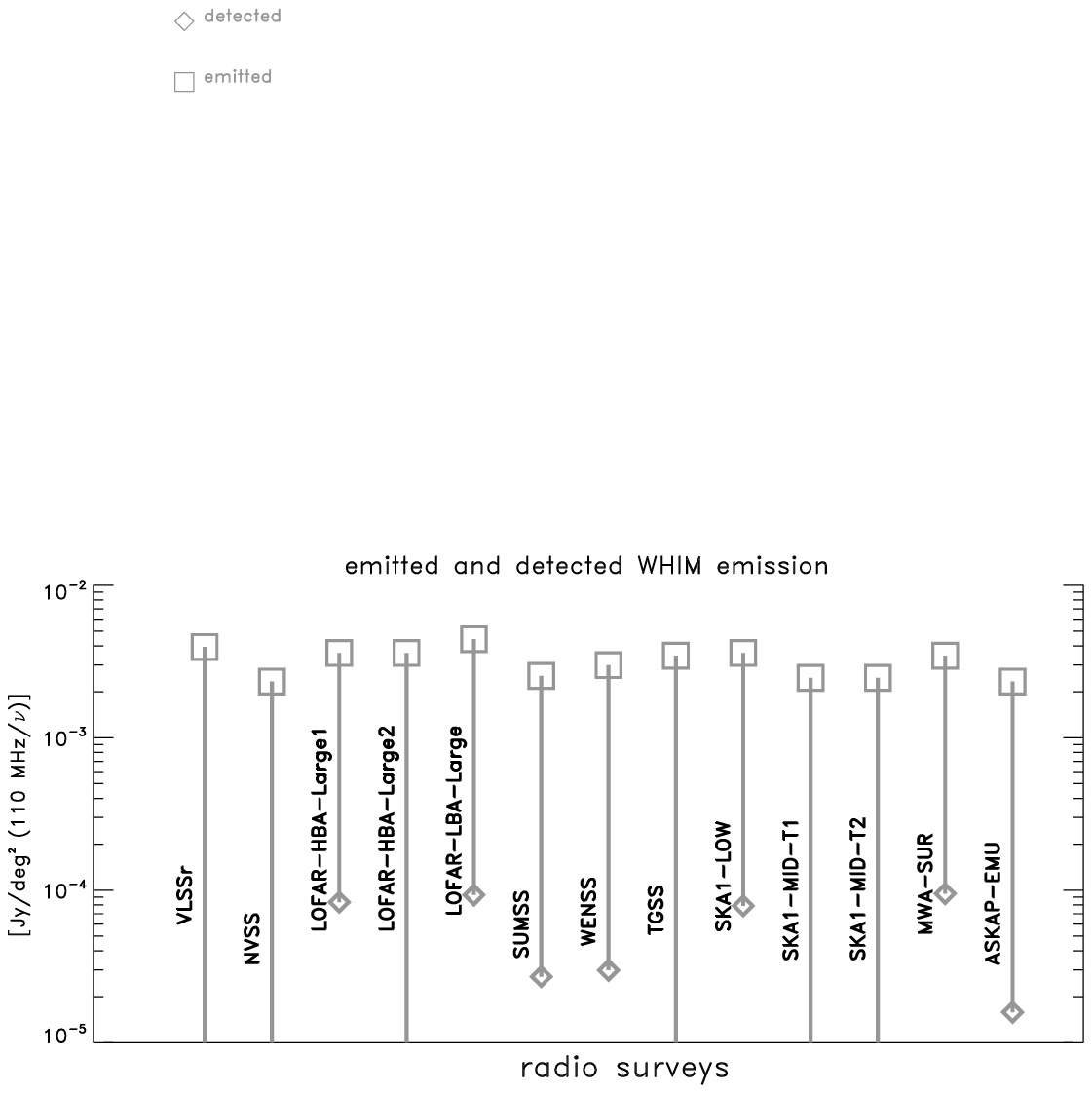}
\caption{Performances of different radio surveys on a  $14^{\circ} \times 14^{\circ}$ radio sky. The bold squares show the intrinsic emission from the WHIM in the simulated volume, the connected small diamonds show the detectable emission by each different radio survey. To better compare the surveys at different frequencies, all emissions have been rescaled by $(110 ~\rm MHz/\nu)$.}
\label{fig:survey}
\end{figure*}

Figure \ref{fig:survey} summarises the level of radio emission from the shocked WHIM phase within a  projected area of $14^{\circ} \times 14^{\circ}$ in the sky, observed with 13 different radio surveys (for the specific parameters, see Table 2 in \citealt{va15radio}). The smaller diamonds show the detectable emission by each different radio survey after post-processing for its specific parameters and assuming detection at $\geq 3 \sigma_{\rm rms}$. Most of the detectable radio cosmic web is confined to $z \leq 0.1-0.2$, because of cosmological dimming. The largest fraction of detected flux from the WHIM is expected by LOFAR (Tier 1 surveys in HBA and LBA), MWA (Broadband survey) and SKA-LOW (already during Phase 1). However, on average only a few percent of the total WHIM emission will be detected even in this case. We estimate a background of unresolved radio emission from the radio cosmic web of $\sim 10^{-3}-10^{-2} \rm Jy/deg^2 (100 MHz/\nu)$, in case of high magnetisation of the WHIM.

\subsection{Latest results}
\label{latest}

With newest simulations we are addressing the capabilities of low-frequency radio survey to distinguish among competing magnetisation scenarios for the large-scale structures.
The top panels in Figure 2 show the mass-weighted mean magnetic field along the line of sight through the primordial (left) and astrophysical (right) models introduced in Sec.\ref{models}. 
While the two scenarios by design produce similar average magnetic field values in halos ($\sim 0.1-1 \rm ~\mu G$ as a volume average), they display a very different radial trend moving towards clusters and voids, with the primordial scenario producing a typically higher magnetisation in the filaments. This is expected, because the magnetisation of filaments is mostly driven by compression and carries memory of the initial seed, which is larger in the primordial run.  We notice that the magnetic field seed employed here is among the largest it is allowed by the latest constraints from Planck satellite \citep[][]{Planck15}.  In the astrophysical scenario instead, filaments are characterised by a more irregular distribution of magnetic fields, as an effect of the extra compression by radiative cooling and by the release of extra magnetic fields close to AGN. 
Away from these two sources of amplification, the field is on average $\sim 100$ lower that in the primordial scenario. 

The central panels of Fig.2 show the radio emission in the two scenarios, computed at $120 ~\rm MHz$ and   assuming the emission to be located at $z=0.025$ ($d_L \approx 100 ~\rm Mpc$).  The lower panels of Fig.2 show that even with a LOFAR-HBA Tier 1 survey, with sensitivity $\sigma_{\rm rms} \approx 0.25 ~\rm mJy/beam$ (corresponding to $\approx 0.35 ~\rm \mu Jy/arcsec^2$ for a resolution beam of of $25$ arcsec) it will be hard to detect the WHIM emission, consistent with our previous findings (Sec.\ref{old}).
Apart from the  radio-relic like emission within clusters/groups and due to internal merger shocks, only a few localised patches in the cosmic web emits above the $\sim ~\rm \mu Jy/arcsec^2$ level, necessary for a $\geq 3 \sigma_{\rm rms}$ detection. 
Radio observations at this rather low level of detection should explore statistical techniques to explore the magnetic cosmic web, like correlating with large-scale structures and stacking multiple observations \citep[e.g.][]{2011JApA...32..577B}.
To better characterise the intrinsic difference in the scale distribution of emission in the two scenarios we computed the 2-point correlation function of radio emission in the two mock observations (Fig.3). The correlation function is computed as the average of all $C(l)=\langle I(r) \cdot I(r+l))\rangle$ (where $I(r)$ is the
radio flux at a given position in the maps) between all detectable pixels in the mock observation. In Fig.3 we show the result of assuming a noise of  $\sigma_{\rm rms}=0.3 ~\rm \mu Jy/arcsec^2$ as well as a ten times smaller one,  $0.03 ~\rm \mu Jy/arcsec^2$. 
In both cases, the difference between the primordial and the astrophysical scenario is very significant for $l \leq 10 \rm Mpc$, corresponding to $\sim 5.5^{\circ}$ in the sky at this redshift.  This follows from the larger magnetisation level of filaments, which creates patches of enhanced emission on scales of several $\sim ~\rm Mpc$ away from clusters. 
However, the fraction of detected pixels is small in the LOFAR-HBA Tier 1 configuration assumed here, and the measured correlation function is noisy, indicating that averaging over larger areas will be fundamental to recover a better signal. In the primordial scenario the correlation
function becomes significantly less noisy with an improved sensitivity, while the gain is less significant in the astrophysical scenario as most of filaments remain too dim to be detected even in this case. 
Within the simple approximations assumed here (removal of galaxy contribution, removal of the galactic foreground, etc), the statistical analysis of large areas of the sky surveyed at low radio frequency promises to be an important tool to discriminate 
among competing scenarios for the origin of extragalactic  magnetic fields. 


\begin{figure*}
\label{fig1}
 \includegraphics[width=0.99\textwidth]{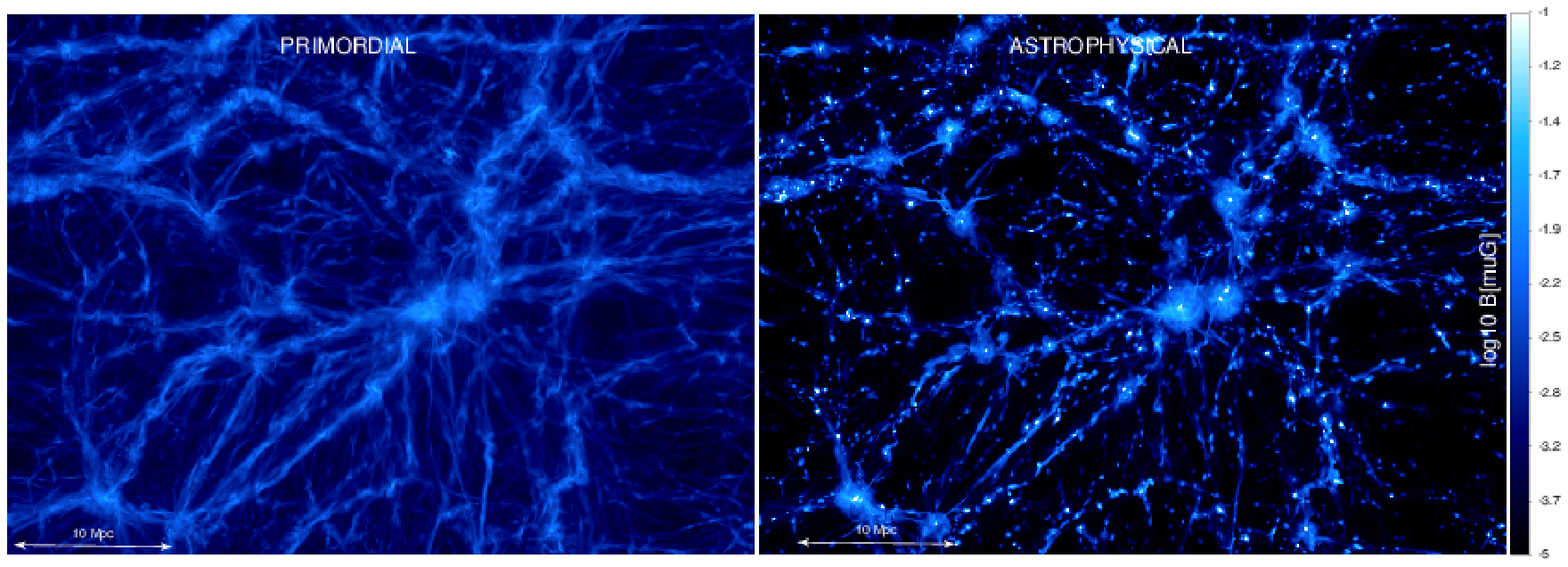}
 \includegraphics[width=0.99\textwidth]{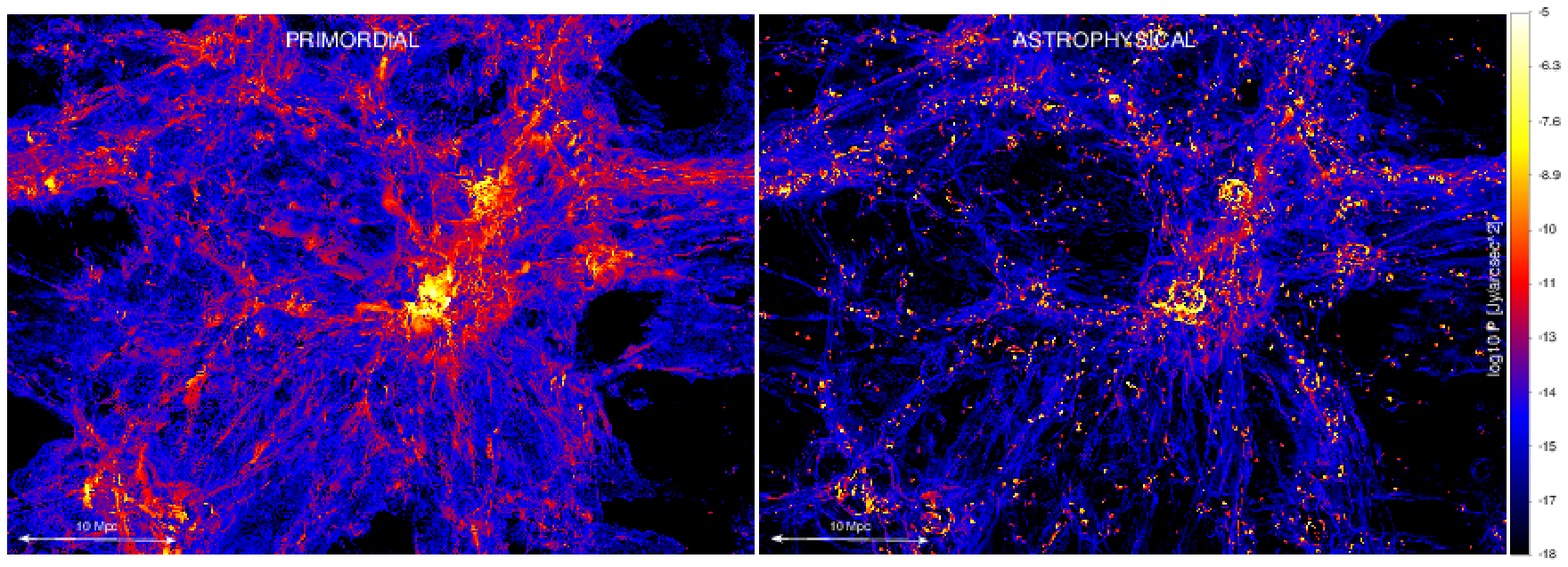}
 \includegraphics[width=0.99\textwidth]{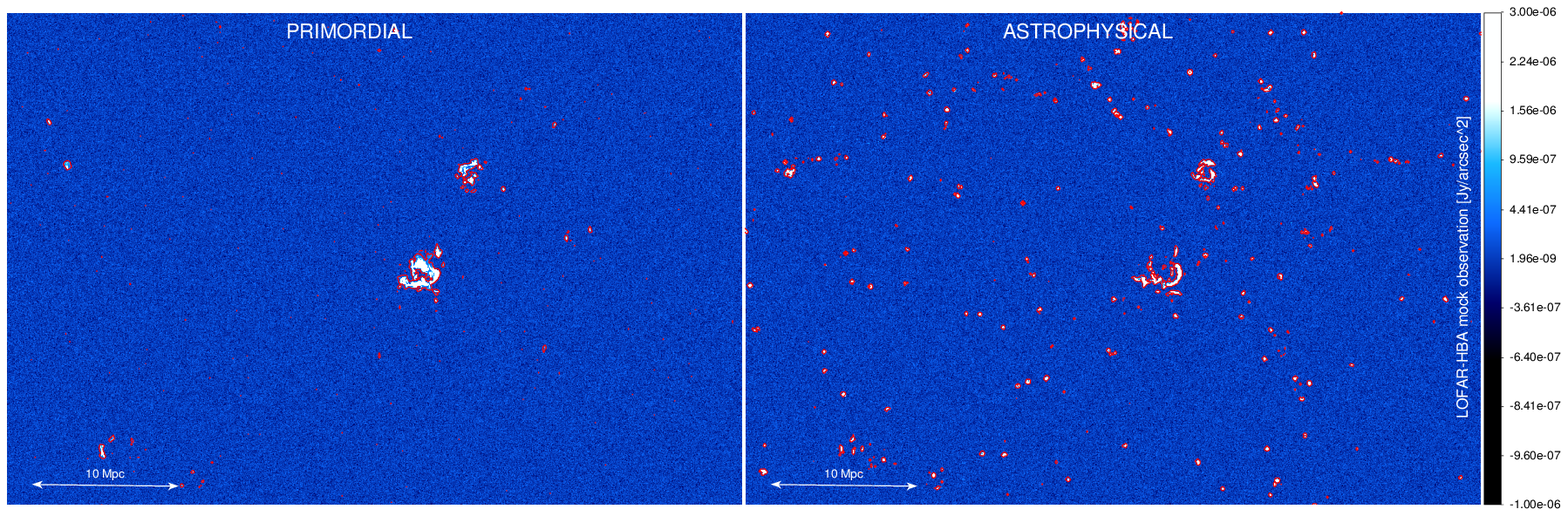}
\caption{Top: mass-weighted mean magnetic fields for a simulated volume ($z=0.025$) for the primordial and the astrophysical magnetisation scenarios. Center: radio emission at $120 ~\rm MHz$ from shock accelerated electrons in the two cases. Bottom: simulated observation with LOFAR-HBA  at $120 ~\rm MHz$. The red contours show the emission above the assumed thermal noise ($\sigma_{\rm rms}=0.3 ~\rm \mu Jy/arcsec^2$).}
\end{figure*}

\begin{figure*}
\label{fig2}
 \includegraphics[width=0.8\textwidth]{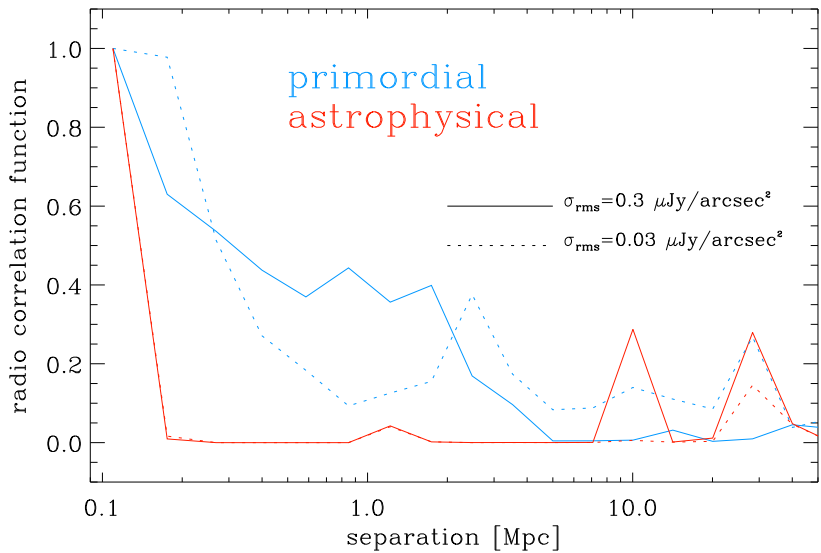}
\caption{Two-point correlation function for detectable pixels in the maps of Fig.2, assuming a noise of $\sigma_{\rm rms}=0.3 ~\rm \mu Jy/arcsec^2$ or $0.03 ~\rm \mu Jy/arcsec^2$.}
\end{figure*}

\section{Conclusions}

The radio cosmic web may be detectable with the incoming generation of radio surveys operating at low frequencies ($\leq 200 ~\rm MHz$). 
Using MHD cosmological simulations, we 
studied the detectability of the shocked and magnetised cosmic underling realistic surveying performances. The telescopes with the best chances of detection are LOFAR-HBA and LBA, SKA1-LOW and MWA, owing to their high sensitivity and better sampling of the small baselines, which allows a better sampling of the large-scale, low surface brightness of the cosmic web. \\
Once removed from the contamination by foregrounds and radio galaxies,  the radio emission level from the cosmic web will enable to constrain the average magnetisation level of the WHIM and the acceleration efficiency of electrons by strong shocks, $\mathcal{M} \geq 10$ \citep{va15ska,va15radio}.\\
With more advanced simulations including the additional magnetisation by active galactic nuclei we showed that statistical methods (i.e. 2-point correlation functions  of the detected radio emission) will have the potential to distinguish among competing scenarios for the magnetisation of large-scale structures in the local universe, provided that electrons are efficiently accelerated by shocks and that some fraction of the cosmic web is bright enough to be detectable.  
Together with the development of the robust statistical tool analyse the typically low signal to noise ratio expected for these 
observations, the production of sophisticated MHD cosmological simulations is a mandatory step to predict the statistical imprints of
competing scenarios on the radio sky, and enable radio surveys to become efficient probes  of cosmic magnetogenesis.

\section*{Acknowledgements}
This work was strongly supported by computing resources from the Swiss National Supercomputing Centre (CSCS) under projects ID ch2 and s585. FV acknowledges personal support from the grant VA 876/3-1 from the Deutsche Forschungsgemeinschaft. FV and MB also acknowledge support from the grant FOR1254 from the Deutsche Forschungsgemeinschaft. We acknowledge allocations no. 9016 and 9059 on supercomputers at the NIC of the Forschungszentrum J\"{u}lich. 
Computations described in this work were performed using the {\enzo} code (http://enzo-project.org), which is the product of a collaborative effort of scientists at many universities and national laboratories.

\bibliographystyle{apj}
\bibliography{franco}

\begin{thebibliography}{}
\expandafter\ifx\csname natexlab\endcsname\relax\def\natexlab#1{#1}\fi

\bibitem[{{Blandford} \& {Eichler}(1987)}]{be87}
{Blandford}, R., \& {Eichler}, D. 1987, \physrep, 154, 1

\bibitem[{{Bonafede} {et~al.}(2013){Bonafede}, {Vazza}, {Br{\"u}ggen},
  {Murgia}, {Govoni}, {Feretti}, {Giovannini}, \& {Ogrean}}]{bo13}
{Bonafede}, A., {Vazza}, F., {Br{\"u}ggen}, M., {et~al.} 2013, \mnras, 433,
  3208

\bibitem[{{Brown}(2011)}]{2011JApA...32..577B}
{Brown}, S.~D. 2011, Journal of Astrophysics and Astronomy, 32, 577

\bibitem[{{Bryan} {et~al.}(2014){Bryan}, {Norman}, {O'Shea}, {Abel}, {Wise},
  {Turk}, {Reynolds}, {Collins}, {Wang}, {Skillman}, {Smith}, {Harkness},
  {Bordner}, {Kim}, {Kuhlen}, {Xu}, {Goldbaum}, {Hummels}, {Kritsuk}, {Tasker},
  {Skory}, {Simpson}, {Hahn}, {Oishi}, {So}, {Zhao}, {Cen}, {Li}, \& {Enzo
  Collaboration}}]{enzo14}
{Bryan}, G.~L., {Norman}, M.~L., {O'Shea}, B.~W., {et~al.} 2014, \apjs, 211, 19

\bibitem[{{Cho}(2014)}]{ch14}
{Cho}, J. 2014, \apj, 797, 133

\bibitem[{{Dav{\'e}} {et~al.}(2001){Dav{\'e}}, {Cen}, {Ostriker}, {Bryan},
  {Hernquist}, {Katz}, {Weinberg}, {Norman}, \& {O'Shea}}]{2001ApJ...552..473D}
{Dav{\'e}}, R., {Cen}, R., {Ostriker}, J.~P., {et~al.} 2001, \apj, 552, 473

\bibitem[{{Donnert} {et~al.}(2009){Donnert}, {Dolag}, {Lesch}, \&
  {M{\"u}ller}}]{donn09}
{Donnert}, J., {Dolag}, K., {Lesch}, H., \& {M{\"u}ller}, E. 2009, \mnras, 392,
  1008

\bibitem[{{Gheller} {et~al.}(2015){Gheller}, {Vazza}, {Favre}, \&
  {Br{\"u}ggen}}]{gheller15}
{Gheller}, C., {Vazza}, F., {Favre}, J., \& {Br{\"u}ggen}, M. 2015, \mnras,
  453, 1164

\bibitem[{{Hoeft} \& {Br{\"u}ggen}(2007)}]{hb07}
{Hoeft}, M., \& {Br{\"u}ggen}, M. 2007, \mnras, 375, 77

\bibitem[{{Keshet} {et~al.}(2004){Keshet}, {Waxman}, \&
  {Loeb}}]{2004ApJ...617..281K}
{Keshet}, U., {Waxman}, E., \& {Loeb}, A. 2004, \apj, 617, 281

\bibitem[{{Marinacci} {et~al.}(2015){Marinacci}, {Vogelsberger}, {Mocz}, \&
  {Pakmor}}]{2015MNRAS.453.3999M}
{Marinacci}, F., {Vogelsberger}, M., {Mocz}, P., \& {Pakmor}, R. 2015, \mnras,
  453, 3999

\bibitem[{{Mogavero} \& {Schekochihin}(2014)}]{2014MNRAS.440.3226M}
{Mogavero}, F., \& {Schekochihin}, A.~A. 2014, \mnras, 440, 3226

\bibitem[{{Pfrommer} {et~al.}(2006){Pfrommer}, {Springel}, {En{\ss}lin}, \&
  {Jubelgas}}]{pf06}
{Pfrommer}, C., {Springel}, V., {En{\ss}lin}, T.~A., \& {Jubelgas}, M. 2006,
  \mnras, 367, 113

\bibitem[{{Planck Collaboration} {et~al.}(2015){Planck Collaboration}, {Ade},
  {Aghanim}, {Arnaud}, {Arroja}, {Ashdown}, {Aumont}, {Baccigalupi},
  {Ballardini}, {Banday}, \& et~al.}]{Planck15}
{Planck Collaboration}, {Ade}, P.~A.~R., {Aghanim}, N., {et~al.} 2015, ArXiv
  e-prints, arXiv:1502.01594

\bibitem[{{Ryu} {et~al.}(2003){Ryu}, {Kang}, {Hallman}, \& {Jones}}]{ry03}
{Ryu}, D., {Kang}, H., {Hallman}, E., \& {Jones}, T.~W. 2003, \apj, 593, 599

\bibitem[{{Skillman} {et~al.}(2011){Skillman}, {Hallman}, {O'Shea}, {Burns},
  {Smith}, \& {Turk}}]{sk11}
{Skillman}, S.~W., {Hallman}, E.~J., {O'Shea}, B.~W., {et~al.} 2011, \apj, 735,
  96

\bibitem[{{Sunyaev} \& {Zeldovich}(1972)}]{1972A&A....20..189S}
{Sunyaev}, R.~A., \& {Zeldovich}, Y.~B. 1972, \aap, 20, 189

\bibitem[{{Vazza} {et~al.}(2013){Vazza}, {Br{\"u}ggen}, \&
  {Gheller}}]{va13feedback}
{Vazza}, F., {Br{\"u}ggen}, M., \& {Gheller}, C. 2013, \mnras, 428, 2366

\bibitem[{{Vazza} {et~al.}(2014){Vazza}, {Br{\"u}ggen}, {Gheller}, \&
  {Wang}}]{va14mhd}
{Vazza}, F., {Br{\"u}ggen}, M., {Gheller}, C., \& {Wang}, P. 2014, \mnras, 445,
  3706

\bibitem[{{Vazza} {et~al.}(2009){Vazza}, {Brunetti}, \& {Gheller}}]{va09shocks}
{Vazza}, F., {Brunetti}, G., \& {Gheller}, C. 2009, \mnras, 395, 1333

\bibitem[{{Vazza} {et~al.}(2015{\natexlab{a}}){Vazza}, {Ferrari}, {Bonafede},
  {Br{\"u}ggen}, {Gheller}, {Braun}, \& {Brown}}]{va15ska}
{Vazza}, F., {Ferrari}, C., {Bonafede}, A., {et~al.} 2015{\natexlab{a}},
  Advancing Astrophysics with the Square Kilometre Array (AASKA14), 97

\bibitem[{{Vazza} {et~al.}(2015{\natexlab{b}}){Vazza}, {Ferrari},
  {Br{\"u}ggen}, {Bonafede}, {Gheller}, \& {Wang}}]{va15radio}
{Vazza}, F., {Ferrari}, C., {Br{\"u}ggen}, M., {et~al.} 2015{\natexlab{b}},
  \aap, 580, A119

\bibitem[{{Wang} {et~al.}(2010){Wang}, {Abel}, \& {Kaehler}}]{wang10}
{Wang}, P., {Abel}, T., \& {Kaehler}, R. 2010, \na, 15, 581

\bibitem[{{Widrow} {et~al.}(2012){Widrow}, {Ryu}, {Schleicher}, {Subramanian},
  {Tsagas}, \& {Treumann}}]{wi11}
{Widrow}, L.~M., {Ryu}, D., {Schleicher}, D.~R.~G., {et~al.} 2012, \ssr, 166,
  37

\end{thebibliography}

\end{document}